\documentclass{osa-article}
\usepackage{braket}
\journal{oe}

\articletype{Research Article}

\begin{document}

\title{Generation and measurement of a squeezed vacuum up to 100 MHz at 1550 nm with a semi-monolithic optical parametric oscillator designed towards direct coupling with waveguide modules}

\author{Naoto Takanashi,\authormark{1,2} Wataru Inokuchi,\authormark{1} Takahiro Serikawa,\authormark{1} and Akira Furusawa\authormark{1,*}}

\address{\authormark{1}Department of Applied Physics, School of Engineering, The University of Tokyo, 7-3-1 Hongo, Bunkyo-ku, Tokyo, 113-8656, Japan}
\address{\authormark{2}takanashi@alice.t.u-tokyo.ac.jp} 

\email{\authormark{*}akiraf@ap.t.u-tokyo.ac.jp} 

\homepage{http://www.alice.t.u-tokyo.ac.jp} 


\begin{abstract}
We report generation and measurement of a squeezed vacuum from a semi-monolithic Fabry-P\'erot optical parametric oscillator (OPO) up to 100 MHz at 1550 nm. The output coupler of the OPO is a flat surface of a nonlinear crystal with partially reflecting coating, which enables direct coupling with waveguide modules. Using the OPO, we observed 6.2dB of squeezing at 2 MHz and 3.0 dB of squeezing at 100 MHz. The OPO operates at the optimal wavelength to minimize propagation losses in silica waveguides and looks towards solving a bottleneck of downsizing these experiments: that of coupling between a squeezer and a waveguide.
\end{abstract}

\section{Introduction}
In quantum optics, a quadrature squeezed state has many applications as a resource of non-classicality \cite{Squeeze:Walls}. As a squeezed vacuum is a  phase-sensitive state, it is used for high precision measurements such as gravitational wave detection \cite{interferometer:Squeezing,Gravitationalprototype:Squeezing,Gravitational:Squeezing}. A squeezed vacuum is also a quantum resource in continuous-variable quantum information processing \cite{CVQI:PvL}. Several quantum states such as a Schr\"{o}dinger-cat-like state \cite{Schrodinger:Welsch,Schrodinger:Warit} or a cluster state \cite{Cluster:Squeezing} are created from squeezed vacua. A squeezed vacuum is also essential in quantum information processing such as quantum non-demolition gate \cite{QND:Shiozawa} and one-way quantum computation\cite{Oneway:Yokoyama}.

The first observation of a squeezed vacuum was realized by four-wave mixing in 1985 \cite{fourwaveSq:Valley}. In the following year, a squeezed vacuum was generated using an optical parametric oscillator (OPO) \cite{PDCSq:Kimble}, making use of the cavity-enhancement of the efficiency of parametric process \cite{SqInCav:Yurke}.
Since then, a number of groups started to compete to achieve high-level squeezing \cite{30yeaesSq:Leuchs}, with 7 dB observed at 860 nm using a bow-tie-type OPO in 2006 \cite{7dBSq:Sasaki}, and 15 dB at 1064 nm with a semi-monolithic OPO reported \cite{15dBSq:Schnabel}. This is the best observed squeezing level to date.

In quantum information processing with time-domain multiplexing, a broadband squeezed vacuum is essential to realize fast quantum processing \cite{MillionModes:Yoshikawa}. Although cavity structures offer advantages in the efficiency of the parametric process, it limits the bandwidth of the process with the round-trip. Round-trip optical length of a cavity is inversely proportional to resonance width. Therefore, small cavities are used to obtain broadband squeezed vacua \cite{OPO:Serikawa,BroadbandOPO:Breitenbach,GHzSq:Schnabel}. In 2013, 4.8 dB squeezing from 5 to 100 MHz and 3 dB squeezing from 100 MHz to 1.2 GHz at 1550 nm was achieved using a small double-resonant monolithic OPO with a 2.6 mm long type-0 phase matched KTiOPO${}_{4}$ (PPKTP) crystal \cite{GHzSq:Schnabel}.

In general, there is a trade-off between bandwidth and the pump power required to obtain large squeezing level \cite{theory:OPOthreshold}. 
This is because the effective length of the parametric process decreases as the cavity confinement is weakened or the length of the nonlinear crystal shortened to get a larger bandwidth. Practically, there is also the condition that the intensity of pump beam is low enough not to damage optical elements or the crystal, and so there is a trade-off between bandwidth and squeezing level. In a previous experiment with a monolithic OPO \cite{GHzSq:Schnabel}, 3 dB of squeezing at 1.2 GHz was observed but required internal pump powers as high as 37 W, which was large enough to cause cavity mode deformation.
In another experiment with 9.3 mm long semi-monolithic PPKTP OPO \cite{10mm:OPO}, the oscillation threshold was estimated to be as low as 221 mW but the HWHM linewidth of the cavity was only 21.5 MHz. Since an optical parametric amplifier (OPA) does not have any cavity structure, it allows THz-order bandwidth limited only by dispersion or phase matching conditions \cite{BroadbandOPA,FiberBroadbandOPA,WaveguideBroadbandOPA,OPAFurusawa} however the observed squeezing levels with an OPA are not as high as that with an OPO. Although 5.8 dB of pulsed squeezing at 1064 nm has been observed using high-intensity pulses of light\cite{pulsedOPAsq}, squeezing with continuous wave of light remains at low level such as 2.2 dB at 1064 nm \cite{OPA22dB}.

For large-scale quantum information processing, it is important to realize direct coupling between a source of squeezed vacuum and waveguide modules. In 2015, integration of universal linear optics in a silica chip was performed \cite{UniversalChip:Matsuda}. Integrating linear optics of a quantum optical circuit is a promising approach for realization of a large scale circuit \cite{SPIE:NTT,SPIE:Serikawa}. However, there is a bottleneck of downsizing in the coupling between a squeezer and a waveguide chip. In 2016, generation of a squeezed vacuum using a fiber coupled OPA was demonstrated but the squeezing level was 1.8 dB at 1550 nm \cite{fiberedOPA18dB}. In 2018, an OPO with directly fiber-coupled structure was proposed but measured squeezing level using the OPO was up to 1dB at 1064 nm \cite{Fiber:Fabre}. One of the causes of the low squeezing level could be considered to be that a thin tabular crystal was used in the OPO and the useful length of the crystal was only 80 $\mu$m.

In this paper, we report generation and measurement of a squeezed vacuum at 1550 nm from a semi-monolithic OPO with a 5mm-long cuboid-shaped crystal and a curved mirror, which has capability of direct coupling with a waveguide modules. A previous work \cite{Schafermeier} uses a cavity with a similar structure, but it cannot be directly coupled to a waveguide because of a high reflection coating on a crystal. 1550 nm is one of the conventional wavelengths for optical communication, and is the best wavelength in terms of propagation losses in silica waveguides \cite{OpticalFiber}. We measured 6.2 dB squeezing and 8.5 dB anti-squeezing at 2 MHz, and 3.0 dB squeezing and 3.8 dB anti-squeezing at 100 MHz with a moderate pump power up to 360 mW. To detect the broadband squeezed vacuum with high quantum efficiency (QE), we developed a homemade homodyne detector with a fast trans-impedance amplifier consisting of discrete semiconductors and high-QE InGaAs photodiodes, which has 14 dB signal-to-noise ratio (SNR) at 100 MHz with a 3.5 mW local oscillator beam.
\section{Design of OPO}
\begin{figure}[t]
\centering\includegraphics[width=11.5cm]{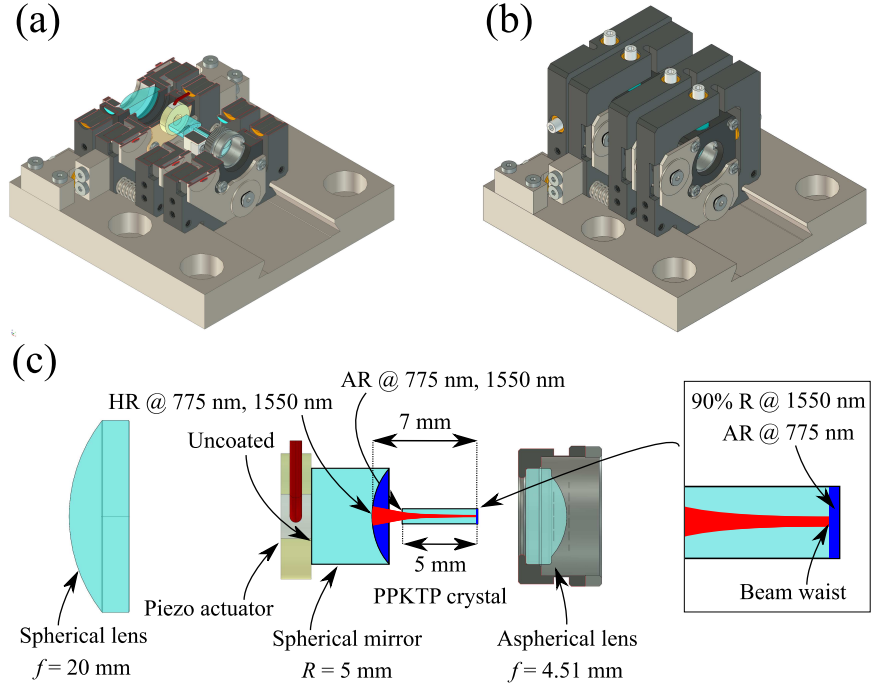}
\caption{Design of our OPO. Optical elements are placed on a dovetail groove. (a) A cross section. (b) An external view. (c) A schematic picture.}
\label{PngOPO}
\end{figure}
Figure \ref{PngOPO} shows the design of our OPO. A Fabry-P\'erot-shaped cavity consists of a spherical mirror (LAYERTEC, curvature radius 5.0 mm, diameter 6.35 mm) and a type-0 phase matched KTiOPO${}_{4}$ (PPKTP) crystal (Raicol Crystals, 1.0 mm $\times$ 1.0 mm $\times$ 5.0 mm) with 90\% reflection coating at 1550 nm on one side. 
The actual physical length of the cavity is 7 mm and taking the refractive index into account, the round-trip optical length is approximately 22 mm. The FWHM linewidth of the cavity is 2.4$\times 10^{2}$ MHz and the finesse of the cavity is 61.

The spherical mirror has a high reflection (HR) coating for both 1550 nm and 775 nm on the spherical surface, and is uncoated on the other side. The spherical mirror is glued to a ring-shaped piezo actuator (Thorlabs, PA44LEW) for cavity length control. A weak coherent beam for cavity locking is injected from the uncoated surface of this mirror. 

The crystal has partial-reflection coating (90\% R at 1550 nm and high transmission (HT) at 775 nm) on one side, and anti-reflection coating for both 1550 nm and 775 nm on the other side. The surface with the partial-reflection coating is the output coupler of the OPO, and a pump beam is also injected from this surface. The waist radius of a resonant beam on the surface of the crystal is 23 $\mu$m.

Focusing lenses are placed either side of the OPO cavity. The one side which the squeezed vacuum exits from has an aspherical lens (Sigma koki, A355T with special ordered anti-reflection coating at both 1550 nm and 775 nm) to obtain better mode-matching with a local oscillator beam for homodyne detection, and the other side has a spherical lens (Thorlabs, LA1074-C).

The beam waist of resonant modes of the OPO is designed to be at a surface of the crystal. Thus, the wavefront is flat on the output coupler, which allows direct coupling with waveguide modules. In the OPO, every optical element is placed on a dovetail groove and has three degrees of freedom of translation. Providing the OPO with a mechanism for adjusting the position of the resonant mode could be useful for coupling several OPOs to one optical integrated chip directly.

Because of the high reflectivity of the spherical mirror at 775 nm, the crystal is pumped from both sides. Since there is non-zero phase shift on reflection on the mirror and in propagation at the boundaries of the periodically poled crystal, using both forward and backward paths is not equivalent to doubling the length of the crystal. In this case, phase matching condition becomes more complex. Here, we discuss the phase matching condition with an additional phase shift $\theta$ between forward and backward paths using second harmonic generation (SHG) as an example.

The total second harmonic output is a superposition of frequency doubled beams from
the forward and backward paths
\begin{equation}
  E_{\scalebox{0.75}{total}} = E_{\scalebox{0.75}{forward}} + E_{\scalebox{0.75}{backward}},
\end{equation}
where $E_{\scalebox{0.75}{forward}}$ is the frequency doubled beam from the forward path, and $E_{\scalebox{0.75}{backward}}$ is that from the backward path. The frequency doubled beams from a periodically poled crystal can be written as \cite{FirstQuasiPM,TheoryOfQPM}:
\begin{equation}
  E_{\scalebox{0.75}{forward}} = AL \mbox{sinc} \left( \frac{\Delta k_{Q}L}{2} \right)
\end{equation}
\begin{equation}
  E_{\scalebox{0.75}{backward}} = AL \mbox{sinc} \left( \frac{\Delta k_{Q}L}{2} \right) e^{i\left(\Delta k_{Q}L+\theta\right)},
\end{equation}
where $A$ is a constant proportional to intensity of a fundamental beam and effective nonlinear efficiency of the crystal, and $L$ is the length of the crystal, $\Delta k_Q = 2\pi\left(\frac{n_{2f}}{\lambda_{2f}}-2\frac{n_{f}}{\lambda_{f}}-\frac{1}{\Lambda}\right)$, and $\Lambda$ is the period of the polarization.

The total intensity is
\begin{equation}
  |E_{\scalebox{0.75}{total}}|^2 = (A\cdot2L)^2 \mbox{sinc}^2 \left( \frac{\Delta k_{Q}L}{2} \right) \cos^2\left( \frac{\Delta k_{Q}L}{2}+\theta \right),
  \label{EquationSHGtotal}
\end{equation}
which has a maximum at $\theta=0{}^{\circ}$
\begin{equation}
  |E_{\scalebox{0.75}{total}}|^2 = (A\cdot2L)^2 \mbox{sinc}^2 \left( \Delta k_{Q}L \right)
\end{equation}
and minimum when $\theta = 90{}^{\circ}$
\begin{equation}
  |E_{\scalebox{0.75}{total}}|^2 = (A\cdot2L)^2 \left( \frac{1-\cos(\Delta k_{Q}L)}{\Delta k_{Q}L} \right)^2.
\end{equation}
The output for $\theta= 90^{\circ}$ is 0.525 times that for $\theta= 0^{\circ}$ but still larger than the individual values of $|E_{\scalebox{0.75}{forward}}|^2$ or $|E_{\scalebox{0.75}{backward}}|^2$.
Experimentally, $\Delta k_{Q}L$ can be scanned by changing the temperature of the crystals, and $\theta$ can be estimated from the ratio of the intensity of the largest peak to that of the second largest peak. In the OPO, $\theta$ is estimated to be $75^{\circ}$. Note that if a pump beam is reflected only on surfaces of a crystal as for example monolithic cavities in previous works \cite{doublepass1,doublepassnot}, this problem is avoided by dicing the crystal at appropriate positions.

The conversion efficiency of SHG is measured to be 2.24 W${}^{-1}$ in the OPO. Taking the cavity enhancement factor ${T^2}/{(1-\sqrt{R})^4}\approx1.44\times10^{3}$ \cite{SHGinCav} into account, non-linear conversion coefficient $E_{NL}$ of the double pass is estimated to be 1.56$\times 10^{-3}$ W${}^{-1}$.

\section{Experimental configuration}
\begin{figure}[t]
\centering\includegraphics[width=11cm]{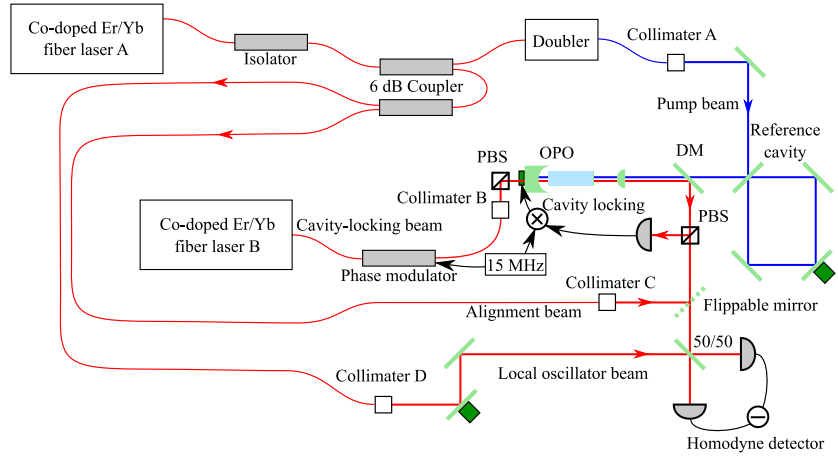}
\caption{Schematic of the experimental setup. Co-doped Er/Yb fiber laser A and B are CW single frequency lasers at 1550 nm. Output of Fiber laser A is split into two beams by means of a 6 dB (75:25) coupler. The brighter beam is frequency doubled by a doubler module and used as the pump beam of the OPO. An alignment beam, a beam for generating a second harmonic in the OPO during alignment, is injected through a flippable mirror when needed. The squeezed vacuum from the OPO get interfered with a local oscillator beam collimated by a triplet lens (Collimater D) in a 50/50 beamsplitter and detected by a homodyne detector. Note that only important elements are depicted. FG, function generator; DM, dichroic mirror; PBS, polarizing beamsplitter; 50/50, 50/50 beamsplitter.}
\label{Setup}
\end{figure}
Figure \ref{Setup} shows a schematic of the experiment. Sources of continuous-wave laser light at 1550 nm are two single frequency Co-doped Erbium/Ytterbium fiber laser systems with different output intensities. Fiber laser A, the brighter one (NKT Photonics, Koheras BOOSTIK C15) is used as the main laser of this experiment with a maximum output power is 2 W. Fiber laser B, the less bright one (NKT Photonics, Koheras ADJUSTIK C15) is used as a source of a reference beam for cavity locking and has a maximum output power of 10 mW. Both lasers are frequency stabilized by slow thermal control and fast piezo control. The output of Fiber laser A is split by a 6dB fiber coupler (Thorlabs, PN1550R3A1) after isolation using a pigtailed isolator (IO-J-1550APC). The main output of the 6 dB fiber coupler pumps a periodically poled lithium niobate waveguide SHG module (NTT Electronics, WH-0775-000-F-B-C) and the output power at 775 nm is 450 mW at maximum. The tapped output of the 6dB fiber coupler is used as a local oscillator (LO) of our homodyne measurement and a reference beam for alignment. Every beam except for the beam for cavity locking is p-polarized.

The OPO is locked by the Pound-Drever-Hall technique \cite{PDH:Drever} with the s-polarized cavity-locking beam. The wavelength of Fiber laser B is set to compensate birefringence of the crystal. The cavity-locking beam is modulated in a phase modulator (Thorlabs, LN65S-FC). The frequency of the modulation is set to 15 MHz to lock the broadband OPO. The cavity-locking beam is separated from the squeezed vacuum by means of a polarizing beamsplitter.

To align the transverse mode of the pump beam, a rectangle-shaped reference cavity is placed between the doubler module and the OPO. During the alignment, a flippable mirror is flipped up and the alignment beam is injected into the OPO to generate second harmonic light. By aligning the mirrors and lenses to make both of the pump beam and the second harmonic light resonate with the reference cavity, the transverse mode of the pump beam is optimized. After the alignment, the optical path in the rectangle-shaped cavity is blocked.

The squeezed vacuum from the OPO is separated from the pump beam by means of a dichroic mirror (Thorlabs, DMSP1180) and gets interfered with the LO beam by a 50/50 beamsplitter (Sigma koki, PSMHQ-25.4C05-10-1550p). The LO beam from an optical fiber is collimated by a triplet lens collimator (Thorlabs, TC06APC) to get better circularity, which provides a visibility of 99\% in the homodyne detection. A mirror glued to a piezo actuator (Thorlabs, AE0505D08F) is placed in the LO path, which is used to scan the phase of the homodyne detection.

The photodiodes of our homemade homodyne detector are specially ordered InGaAs photodiodes (Laser Components, IGHQEX0100-1550-10-1.0-SPAR-TH-40). The trans-impedance amplifier consists of a cascade amplifier (Infineon Technologies, NE3509 and BFR740) and an emitter follower (Infineon Technologies, BFR740). The output electric signal is measured by a spectrum analyzer (Agilent, E4401B).
\section{Results and discussions}
\begin{figure}[t]
\centering\includegraphics[width=11cm]{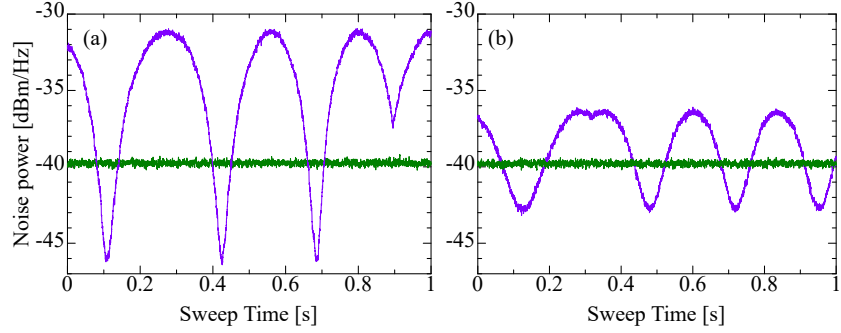}
\caption{Raw data of noise power as a function of the phase of the LO beam (scanned by a triangle wave). Green lines are shot noises (without the pump beam), and purple lines are noises of squeezed vacuum. (a) Center frequency is set to 2 MHz. (b) Center frequency is set to 100 MHz. Resolution bandwidth is set to 1 MHz and video bandwidth is set to 300 Hz. Intensity of the pump beam before the OPO is 360 mW and intensity of the LO beam before the 50/50 beamsplitter is 3.5 mW.}
\label{SqueezingScan_with_360mW}
\end{figure}
\begin{figure}[ht]
\centering\includegraphics[width=9cm]{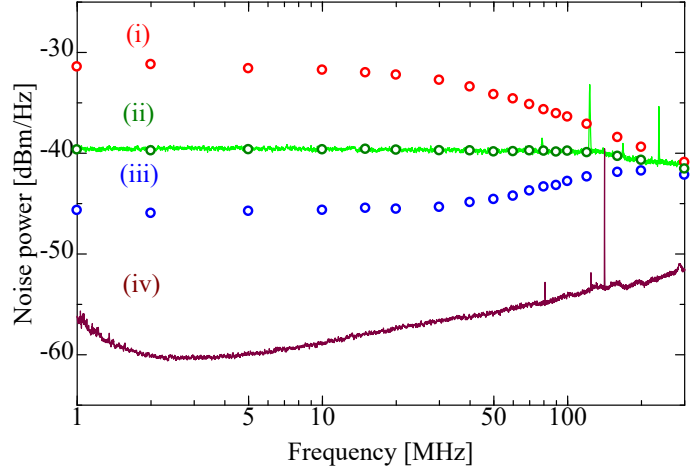}
\caption{Measured noise spectrum from 1 MHz to 300 MHz. (i) Anti-squeezed noise. (ii) Shot noise. (iii) Squeezed noise. (iv) Circuit noise. Each circle is obtained by scanning of the phase of the LO beam as well as Fig. \ref{SqueezingScan_with_360mW}. Lines are raw data from the spectrum analyzer. Every value is without any noise compensation. Resolution bandwidth is set to 1 MHz and video bandwidth is set to 300 Hz. Intensity of the pump beam before the OPO is 360 mW and intensity of the LO beam before the 50/50 beamsplitter is 3.5 mW. Because of a high pass filter with a cutoff frequency of 100 kHz in the homodyne detector, measured points at 1 MHz are slightly distorted.}
\label{SqueezingSpectrum_with_360mW}
\end{figure}
\begin{figure}[!ht]
\centering\includegraphics[width=11cm]{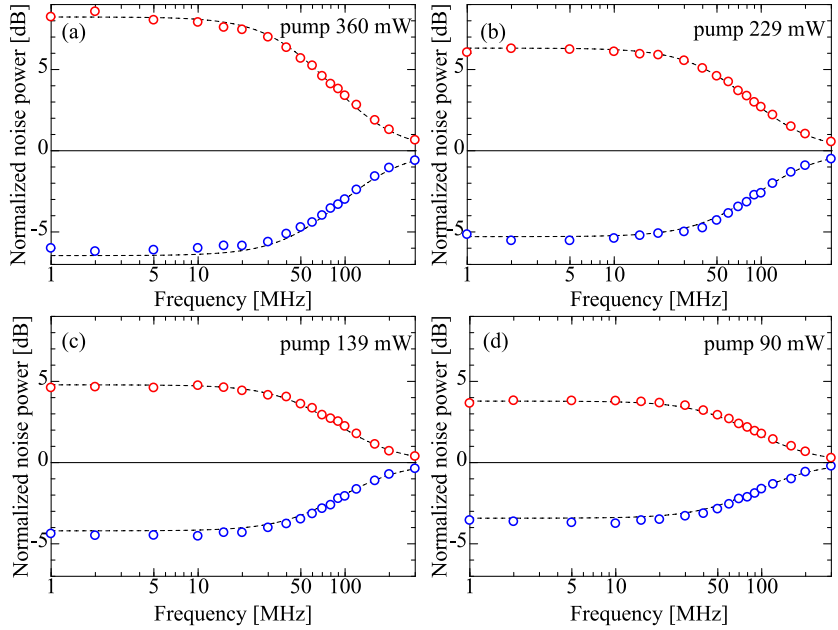}
\caption{Pump power dependence of squeezing level and anti-squeezing level from 1 MHz to 300 MHz. Intensity of the pump beam is (a) 360 mW, (b) 229 mW, (c) 139 mW, (d) 90 mW. Oscillation threshold and $f_{\protect\scalebox{0.5}{HWHM}}$ are fitted to be 1.7 W and 92 MHz. Dashed lines in each graph are a theoretical fitting from Eq. (\ref{SqueezingTheory}).}
\label{NormalizedSpec}
\end{figure}
\begin{figure}[!ht]
\centering\includegraphics[width=11cm]{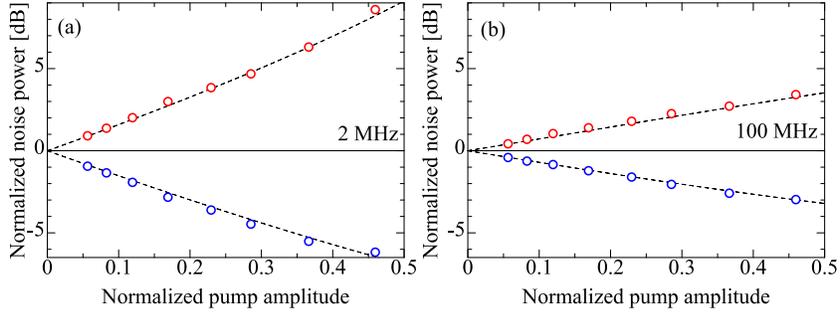}
\caption{Normalized noise power as a function of normalized pump amplitude $\sqrt{\xi}$ (square root of pump power divided by oscillation threshold 1.7 W). (a) at 2 MHz and (b) at 100 MHz. Dashed lines in each graph are a theoretical fitting from Eq. (\ref{SqueezingTheory}).}
\label{PumpDep}
\end{figure}
\begin{figure}[t]
\centering\includegraphics[width=11cm]{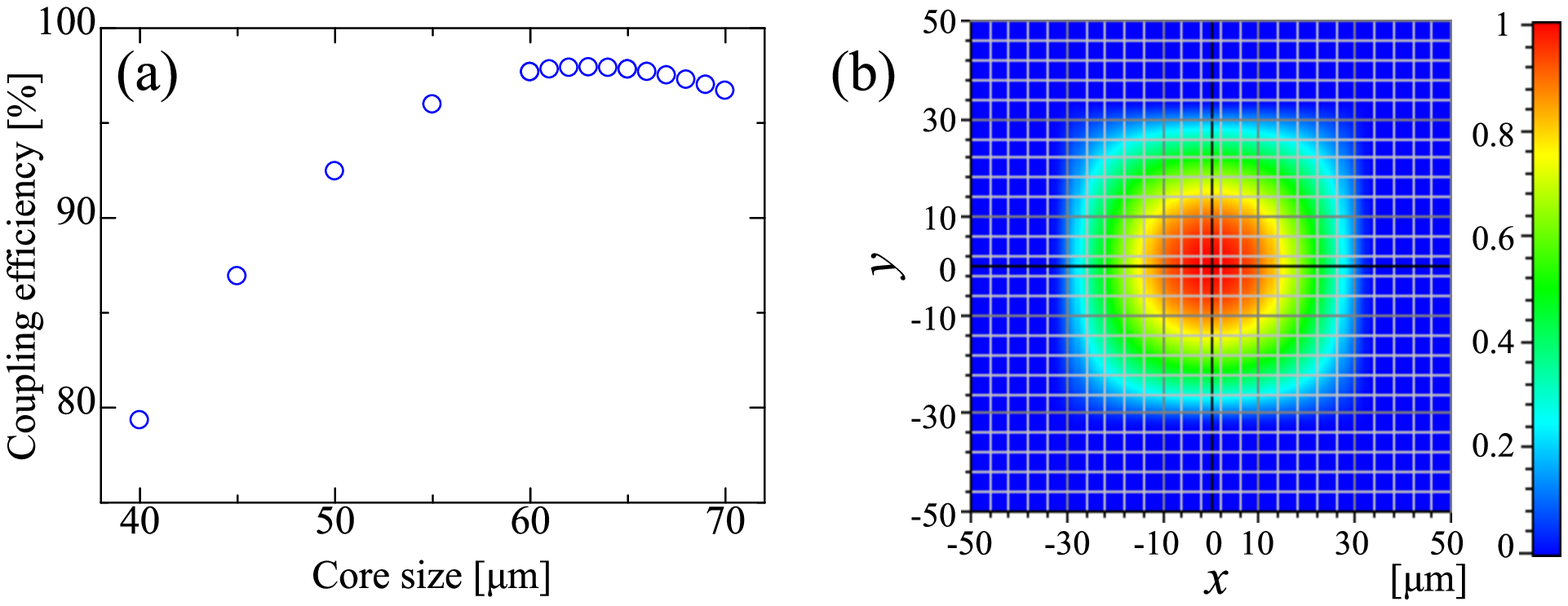}
\caption{Result of simulation. (a) Expected coupling efficiency between the OPO and a square waveguide with each core size. (b) The electric field intensity distribution of the propagation mode in a waveguide with a core size of 63 $\mu$m.}
\label{Simulation}
\end{figure}
Figure \ref{SqueezingScan_with_360mW} shows raw data from the spectrum analyzer at 2 MHz and 100 MHz with 360 mW pump power and 3.5 mW LO power. Observed squeezing level is 6.2$\pm$0.1 dB at 2 MHz and 3.0$\pm$0.1 dB at 100 MHz, and observed anti-squeezing level is 8.6$\pm$0.1 dB at 2 MHz and 3.4$\pm$0.1 dB at 100 MHz. Figure \ref{SqueezingSpectrum_with_360mW} shows observed noise power of the squeezed vacuum at each frequency, the shot noise and the circuit noise. Figures \ref{NormalizedSpec} and \ref{PumpDep} show pump amplitude dependence of the squeezing level and the anti-squeezing level.

In theory, the squeezing level and anti-squeezing level are written as \cite{LossSq:PKLam}: 
\begin{equation}
V_{\pm}(f) = 1 \pm (1-\eta)\frac{T}{T+L}\frac{4\sqrt{\xi}}{(1\mp\sqrt{\xi})^{2}+(f/f_{\scalebox{0.5}{HWHM}})^{2}},
\label{SqueezingTheory}
\end{equation}
where $T$ is the transmittance of the output coupler of the OPO, $L$ is the internal-cavity loss, $\eta$ is total detection loss, $\xi$ is the pump power normalized by oscillation threshold $P_{th}$ of the OPO, and $f$ is the frequency of measurement. In the OPO, measured intra-cavity loss $L$ is 0.38\%, which is considered to be mainly due to imperfections in the anti-reflection coating of the crystal, and the transmittance of the output coupler $T$ is 10\%. The result of the measurement is well-fit by the theoretical formula. We estimate an $\eta$ of 7\% with a break down as follows: 3\% propagation loss; 2\% mode mismatch at the 50/50 beamsplitter of the homodyne measurement; 1\% detection loss at the photodiodes; 1\% equivalent loss of the circuit noise. The oscillation threshold and $f_{\scalebox{0.5}{HWHM}}$ are fitted to be 1.7 W and 92 MHz. The oscillation threshold is well-matched with the estimation from a formula $P_{th}=(T+L)^2 / 4E_{NL}=1.73$ W \cite{7dBSq:Sasaki}.

From Eq. (\ref{EquationSHGtotal}), $E_{NL}$ is estimated to be 0.64 times that of the case of $\theta=0$. Thus, it can be expected that the oscillation threshold can be reduced to 1.1W by optimizing the phase of the reflected pump beam. A way to realize this condition is replacing the spherical mirror (HR at 1550 nm and 775 nm) in the OPO with a dichroic spherical mirror (HR at 1550 nm, AR at 775 nm), and placing a mirror for pump beam at an appropriate position behind the dichroic spherical mirror. Another way to realize the condition is by changing the dicing positions of the periodically poled crystal based on phase shift on the reflection on the spherical mirror.

When the OPO is directly coupled with waveguide modules in the future, pump power could be limited by a damage thresold of the modules. Since the intensity of the squeezed vacuum is generally very low, it is sufficient to consider the dichroic beam splitter, which is the only module in which pump beam propagates in our scheme. A typical maximum power of fiber optic dichroic beamsplitters (such as Thorlabs, WP9850B, WP9864B) is 1W. Taking the reflected pump beam into account, the maximum input pump power is half of that, namely, 0.5 W. In our experiment, the maximum pump power is 360 mW, which is limited by the output power of the laser, and this is still lower than 0.5 W. 

The expected coupling efficiency with waveguide modes was calculated by simulation using a finite-difference method (Optiwave Systems Inc., OptiBPM 12). In anticipation of implementation on a silica-based planar light circuit (PLC) generally used for communication \cite{GeneralPLC150}, the shape of the waveguide is set to be square, and its refractive index difference is set to be 1.5\%. Figure \ref{Simulation} shows the result of the simulation. Ignoring misalignment and reflection, a coupling efficiency of 97.9\% is expected for a linear waveguide with a core size of 63 $\mu$m on a side. The mismatch is considered to be mainly due to the waveguide mode being slightly different from circular. Additionally, in silicon waveguides, efficient spot size conversion is performed using a three-dimensional taper \cite{3Dsilicon}. A tilt dry-etching technique \cite{3DPLC} is considered as a method of making a similar structure in a silica-based PLC.

The following could be considered as potential improvements: Reduction of the required intensity of an incident pump beam by reflecting the pump beam at both ends of the cavity, elimination of the second laser by using cascaded acousto-optic modulators to shift the frequency of a beam.
\section{Conclusion}
Using a semi-monolithic Fabry-P\'erot optical parametric oscillator (OPO) with capability of direct coupling with waveguide modules, we achieved 6.2 dB of squeezing at 2 MHz and 3.0 dB of squeezing at 100 MHz with 360 mW of pump power and 3.5 mW of LO power. This is the first realization of an OPO with capability of direct coupling with waveguide modules at 1550 nm, which is the best wavelength for silica waveguides in terms of propagation losses. The OPO is expected to contribute a great deal to the downsizing of quantum optical circuits.
\section*{Funding}
Core Research for Evolutional Science and Technology (CREST) (JPMJCR15N5) of Japan Science and Technology Agency (JST); KAKENHI (17H01150) of Japan Society for the Promotion of Science (JSPS); APLS of Ministry of Education, Culture, Sports, Science and Technology (MEXT); The University of Tokyo Foundation.
\section*{Acknowledgments}
We are grateful to First Mechanical Design (FMD) Corporation for their assistance with the design and fabrication of the OPO mount system. We thank Euan J. Allen  and Takahiro Kashiwazaki for feedback on the manuscript.
\bibliography{sample}






\end{document}